\documentclass[a4paper,11pt]{article}

\makeatletter
\gdef\@fpheader{   }
\gdef\@journal{jhep}

\RequirePackage{amsmath}
\RequirePackage{amssymb}
\RequirePackage{epsfig}
\RequirePackage{CJK}
\RequirePackage{graphicx}
\RequirePackage[numbers,sort&compress]{natbib}
\RequirePackage{color}
\RequirePackage[
colorlinks=true
,urlcolor=blue
,anchorcolor=blue
,citecolor=blue
,filecolor=blue
,linkcolor=blue
,menucolor=blue
,pagecolor=blue
,linktocpage=true
,pdfproducer=medialab
,pdfa=true
,CJKbookmarks
]{hyperref}

\newif\ifnotoc\notocfalse
\newif\ifemailadd\emailaddfalse
\newif\iftoccontinuous\toccontinuousfalse

\def\@subheader{\@empty}
\def\@keywords{\@empty}
\def\@abstract{\@empty}
\def\@xtum{\@empty}
\def\@dedicated{\@empty}
\def\@arxivnumber{\@empty}
\def\@collaboration{\@empty}
\def\@collaborationImg{\@empty}
\def\@proceeding{\@empty}
\def\@preprint{\@empty}

\newcommand{\subheader}[1]{\gdef\@subheader{#1}}
\newcommand{\keywords}[1]{\if!\@keywords!\gdef\@keywords{#1}\else%
\PackageWarningNoLine{\jname}{Keywords already defined.\MessageBreak Ignoring last definition.}\fi}
\renewcommand{\abstract}[1]{\gdef\@abstract{#1}}
\newcommand{\dedicated}[1]{\gdef\@dedicated{#1}}
\newcommand{\arxivnumber}[1]{\gdef\@arxivnumber{#1}}
\newcommand{\proceeding}[1]{\gdef\@proceeding{#1}}
\newcommand{\xtumfont}[1]{\textsc{#1}}
\newcommand{\correctionref}[3]{\gdef\@xtum{\xtumfont{#1} \href{#2}{#3}}}
\newcommand\jname{JHEP}
\newcommand\acknowledgments{\section*{Acknowledgments}}

\newcommand\preprint[1]{\gdef\@preprint{\hfill #1}}
\newenvironment{proof}[1][Proof]{\noindent\textbf{#1.} }{\ \rule{0.5em}{0.5em}}

\newcommand\note[2][]{%
\if!#1!%
\stepcounter{footnote}\footnotetext{#2}%
\else%
{\renewcommand\thefootnote{#1}%
\footnotetext{#2}}%
\fi}

\newtoks\auth@toks
\renewcommand{\author}[2][]{%
  \if!#1!%
    \auth@toks=\expandafter{\the\auth@toks#2\ }%
  \else
    \auth@toks=\expandafter{\the\auth@toks#2$^{#1}$\ }%
  \fi
}

\newtoks\affil@toks\newif\ifaffil\affilfalse
\newcommand{\affiliation}[2][]{%
\affiltrue
  \if!#1!%
    \affil@toks=\expandafter{\the\affil@toks{\item[]#2}}%
  \else
    \affil@toks=\expandafter{\the\affil@toks{\item[$^{#1}$]#2}}%
  \fi
}

\newtoks\email@toks\newcounter{email@counter}%
\newcommand{\emailAdd}[1]{%
\emailaddtrue%
\ifnum\theemail@counter>0\email@toks=\expandafter{\the\email@toks, \@email{#1}}%
\else\email@toks=\expandafter{\the\email@toks\@email{#1}}%
\fi\stepcounter{email@counter}}
\newcommand{\@email}[1]{\href{mailto:#1}{\tt #1}}

\newcommand*\collaboration[1]{\gdef\@collaboration{#1}}
\newcommand*\collaborationImg[2][]{\gdef\@collaborationImg{#2}}

\newcommand\afterLogoSpace{\smallskip}
\newcommand\afterSubheaderSpace{\vskip3pt plus 2pt minus 1pt}
\newcommand\afterProceedingsSpace{\vskip21pt plus0.4fil minus15pt}
\newcommand\afterTitleSpace{\vskip23pt plus0.06fil minus13pt}
\newcommand\afterRuleSpace{\vskip23pt plus0.06fil minus13pt}
\newcommand\afterCollaborationSpace{\vskip3pt plus 2pt minus 1pt}
\newcommand\afterCollaborationImgSpace{\vskip3pt plus 2pt minus 1pt}
\newcommand\afterAuthorSpace{\vskip5pt plus4pt minus4pt}
\newcommand\afterAffiliationSpace{\vskip3pt plus3pt}
\newcommand\afterEmailSpace{\vskip16pt plus9pt minus10pt\filbreak}
\newcommand\afterXtumSpace{\par\bigskip}
\newcommand\afterAbstractSpace{\vskip16pt plus9pt minus13pt}
\newcommand\afterKeywordsSpace{\vskip16pt plus9pt minus13pt}
\newcommand\afterArxivSpace{\vskip3pt plus0.01fil minus10pt}
\newcommand\afterDedicatedSpace{\vskip0pt plus0.01fil}
\newcommand\afterTocSpace{\bigskip\medskip}
\newcommand\afterTocRuleSpace{\bigskip\bigskip}
\newlength{\affiliationsSep}
\newcommand\beforetochook{\pagestyle{myplain}\pagenumbering{roman}}

\DeclareFixedFont\trfont{OT1}{phv}{b}{sc}{11}

\renewcommand\maketitle{
\pagestyle{empty}
\thispagestyle{titlepage}
\setcounter{page}{0}
\noindent{\small\scshape\@fpheader}\@preprint\par
\afterLogoSpace
\if!\@subheader!\else\noindent{\trfont{\@subheader}}\fi
\afterSubheaderSpace
\if!\@proceeding!\else\noindent{\sc\@proceeding}\fi
\afterProceedingsSpace
{\LARGE\flushleft\sffamily\bfseries\@title\par}
\afterTitleSpace
\hrule height 1.5\p@%
\afterRuleSpace
\if!\@collaboration!\else
{\Large\bfseries\sffamily\raggedright\@collaboration}\par
\afterCollaborationSpace
\fi
\if!\@collaborationImg!\else
{\normalsize\bfseries\sffamily\raggedright\@collaborationImg}\par
\afterCollaborationImgSpace
\fi
{\bfseries\raggedright\sffamily\the\auth@toks\par}
\afterAuthorSpace
\ifaffil\begin{list}{}{%
\setlength{\leftmargin}{0.28cm}%
\setlength{\labelsep}{0pt}%
\setlength{\itemsep}{\affiliationsSep}%
\setlength{\topsep}{-\parskip}}
\itshape\small%
\the\affil@toks
\end{list}\fi
\afterAffiliationSpace
\ifemailadd 
\noindent\hspace{0.28cm}\begin{minipage}[l]{.9\textwidth}
\begin{flushleft}
\textit{E-mail:} \the\email@toks
\end{flushleft}
\end{minipage}
\else 
\PackageWarningNoLine{\jname}{E-mails are missing.\MessageBreak Plese use \protect\emailAdd\space macro to provide e-mails.}
\fi
\afterEmailSpace
\if!\@xtum!\else\noindent{\@xtum}\afterXtumSpace\fi
\if!\@abstract!\else\noindent{\renewcommand\baselinestretch{.9}\textsc{Abstract:}}\ \@abstract\afterAbstractSpace\fi
\if!\@keywords!\else\noindent{\textsc{Keywords:}} \@keywords\afterKeywordsSpace\fi
\if!\@arxivnumber!\else\noindent{\textsc{ArXiv ePrint:}} \href{http://arxiv.org/abs/\@arxivnumber}{\@arxivnumber}\afterArxivSpace\fi
\if!\@dedicated!\else\vbox{\small\it\raggedleft\@dedicated}\afterDedicatedSpace\fi
\ifnotoc\else
\iftoccontinuous\else\newpage\fi
\beforetochook\hrule
\tableofcontents
\afterTocSpace
\hrule
\afterTocRuleSpace
\fi
\setcounter{footnote}{0}
\pagestyle{myplain}\pagenumbering{arabic}
} 

\renewcommand{\baselinestretch}{1.1}\normalsize
\divide\@tempdima\baselineskip
\@tempcnta=\@tempdima
\skip\@mpfootins = \skip\footins
\renewcommand{\@dotsep}{10000}

\newcommand\ps@myplain{
\pagenumbering{arabic}
\renewcommand\@oddfoot{\hfill-- \thepage\ --\hfill}
\renewcommand\@oddhead{}}
\let\ps@plain=\ps@myplain

\newcommand\ps@titlepage{\renewcommand\@oddfoot{}\renewcommand\@oddhead{}}

\numberwithin{equation}{section}

\renewcommand\section{\@startsection{section}{1}{\z@}%
                                   {-3.5ex \@plus -1.3ex \@minus -.7ex}%
                                   {2.3ex \@plus.4ex \@minus .4ex}%
                                   {\normalfont\large\bfseries}}
\renewcommand\subsection{\@startsection{subsection}{2}{\z@}%
                                   {-2.3ex\@plus -1ex \@minus -.5ex}%
                                   {1.2ex \@plus .3ex \@minus .3ex}%
                                   {\normalfont\normalsize\bfseries}}
\renewcommand\subsubsection{\@startsection{subsubsection}{3}{\z@}%
                                   {-2.3ex\@plus -1ex \@minus -.5ex}%
                                   {1ex \@plus .2ex \@minus .2ex}%
                                   {\normalfont\normalsize\bfseries}}
\renewcommand\paragraph{\@startsection{paragraph}{4}{\z@}%
                                   {1.75ex \@plus1ex \@minus.2ex}%
                                   {-1em}%
                                   {\normalfont\normalsize\bfseries}}
\renewcommand\subparagraph{\@startsection{subparagraph}{5}{\parindent}%
                                   {1.75ex \@plus1ex \@minus .2ex}%
                                   {-1em}%
                                   {\normalfont\normalsize\bfseries}}

\def\fnum@figure{\textbf{\figurename\nobreakspace\thefigure}}
\def\fnum@table{\textbf{\tablename\nobreakspace\thetable}}

\long\def\@makecaption#1#2{%
  \vskip\abovecaptionskip
  \sbox\@tempboxa{\small #1. #2}%
  \ifdim \wd\@tempboxa >\hsize
    \small #1. #2\par
  \else
    \global \@minipagefalse
    \hb@xt@\hsize{\hfil\box\@tempboxa\hfil}%
  \fi
  \vskip\belowcaptionskip}

\renewenvironment{thebibliography}[1]{%
\begin{oldthebibliography}{#1}%
\small%
\raggedright%
\setlength{\itemsep}{5pt plus 0.2ex minus 0.05ex}%
}%
{%
\end{oldthebibliography}%
}

\makeatother

\usepackage[T1]{fontenc} 
\usepackage{romannum} 
\usepackage{CJK}

\title{{\boldmath A duality in classical and quantum mechanics: General results}}

\author[a,c,1]{Wen-Du Li}\note{liwendu@tjnu.edu.cn.}
\author[b,2]{and Wu-Sheng Dai}\note{daiwusheng@tju.edu.cn.}

\affiliation[a]{College of Physics and Materials Science, Tianjin Normal University, Tianjin 300387, PR China}
\affiliation[b]{Department of Physics, Tianjin University, Tianjin 300350, P.R. China}
\affiliation[c]{Theoretical Physics Division, Chern Institute of Mathematics, Nankai University, Tianjin, 300071, P. R. China}

\abstract{We reveal a duality in classical and quantum mechanics. Dual systems are
related by duality transforms. All mechanical systems that are dual to each
other form a duality family. In a duality family, once a system is solved, all
other potentials are solved by the dual transform. That is, in a duality
family, we only need to solve one system.}

\begin{document} 
\begin{CJK*}{GBK}{song}
\maketitle 

\flushbottom

\section{Introduction}

Duality reveals underlying connections between different systems. It will be
shown that there exist duality families in classical mechanics and quantum
mechanics. The duality of mechanical systems means that two mechanical systems
are related by a duality relation. Once a dynamical equation --- the Newton
equation or the Schr\"{o}dinger equation --- is solved, the solution of its
dual system can be achieved immediately by the duality relation. All
mechanical systems related by the duality relation form a duality family. A
duality family consists of infinite members. Just need to solve one mechanical
system in the family, the solutions of all other members are achieved by the
duality transform.

A special case of the duality discussed in the present paper was found by
Newton himself in his Principia \cite{chandrasekhar1995newton}, known as the
Newton-Hooke duality. The Newton-Hooke duality is the duality between the
Newton gravity potential and the harmonic-oscillator potential. In this paper,
we provide a general duality relation for arbitrary potentials in classical
mechanics and in quantum mechanics.

Duality is an important concept in modern physics, such as the AdS/CFT
duality\ \cite{maldacena1997large,witten1998anti,witten1998anti2,aharony2000large,d2004supersymmetric}%
, the gravity/fluid duality
\cite{bredberg2012navier,hubeny2011fluid,compere2011holographic,hao2015flat,ashok2014forced,compere2012relativistic,pinzani2015towards}%
, the gravoelectric duality \cite{dadhich2002most}. The modern formulation of
the Newton-Hooke duality can be found in Refs.
\cite{arnold1990huygens,needham1993newton,hall2000planetary}. Recently, there
are authors discuss the power law duality \cite{Inomata2021PowerLaw}.

In sections \ref{classical} and \ref{quantum}, we discuss the duality in
classical mechanics and quantum mechanics. Examples are provided, such as the
duality of the power potential and the Poschl-Teller potential. The conclusion
is given in section \ref{Conclusion}.

\section{Duality in classical mechanics \label{classical}}

The equation of motion in classical mechanics is the Newton equation.

\subsection{One-dimensional potential}

In one dimension, the equation of motion of the potential $U\left(  x\right)
$ with the energy $E$ and the equation of motion of the potential $V\left(
\xi\right)  $ with the energy $\mathcal{E}$ are \textit{ }%
\cite{landau1982mechanics}%
\begin{align}
\frac{dt}{dx}  &  =\frac{1}{\sqrt{2\left[  E-U\left(  x\right)  \right]  }%
},\label{Ueq1d}\\
\frac{d\tau}{d\xi}  &  =\frac{1}{\sqrt{2\left[  \mathcal{E}-V\left(
\xi\right)  \right]  }}. \label{Veq1d}%
\end{align}

\textit{Two one-dimensional potentials, }$U\left(  x\right)  $\textit{ and
}$V\left(  \xi\right)  $\textit{, are dual to each other, if }%
\begin{equation}
x^{-2}\left[  U\left(  x\right)  -E\right]  =\xi^{-2}\left[  V\left(
\xi\right)  -\mathcal{E}\right]  \label{DualoneD}%
\end{equation}
\textit{with}%
\begin{equation}
x\leftrightarrow\xi^{\sigma}. \label{rrho1d}%
\end{equation}
\textit{The solution of the potential }$U\left(  r\right)  $\textit{ with the
energy }$E$\textit{ and the solution of its dual potential }$V\left(
\rho\right)  $\textit{ with the energy }$\mathcal{E}$\textit{\ can be obtained
from each other by the replacement of the time:}%
\begin{equation}
t\leftrightarrow\sigma\tau. \label{thetaphi1d}%
\end{equation}
\textit{Here }$\sigma$\textit{ is a constant chosen arbitrarily.}

\begin{proof}
By\textit{ }Eqs. (\ref{Ueq1d}) and (\ref{Veq1d}) we have
\begin{align}
E-U\left(  x\right)   &  =\frac{1}{2}\left(  \frac{dx}{dt}\right)  ^{2},\\
\mathcal{E}-V\left(  \xi\right)   &  =\frac{1}{2}\left(  \frac{d\xi}{d\tau
}\right)  ^{2}.
\end{align}
Substituting into Eq. (\ref{DualoneD}) gives%
\begin{equation}
x^{-2}\left[  \frac{1}{2}\left(  \frac{dx}{dt}\right)  ^{2}\right]  =\xi
^{-2}\left[  \frac{1}{2}\left(  \frac{d\xi}{d\tau}\right)  ^{2}\right]  .
\end{equation}
This gives%
\begin{equation}
\frac{dt}{d\tau}=\frac{d\ln x}{d\ln\xi}.
\end{equation}
Because $t$ and $\tau$ are independent of $x$ and $\xi$, we have
\begin{equation}
\frac{dt}{d\tau}=\frac{d\ln x}{d\ln\xi}=\sigma, \label{Eqttauxxi}%
\end{equation}
where $\sigma$ is an arbitrary constant. Solving Eq. (\ref{Eqttauxxi}) gives
the duality transformations (\ref{rrho1d}) and (\ref{thetaphi1d}).
\end{proof}

\subsection{Three-dimensional central potential}

In three dimensions, the orbit equation of the central potential $U\left(
r\right)  $ with the energy $E$ and the orbit equation of the central
potential $V\left(  \rho\right)  $ with the energy $\mathcal{E}$ are
\begin{align}
\frac{d\theta}{dr}  &  =\frac{l/r^{2}}{\sqrt{2\left[  E-l^{2}/\left(
2r^{2}\right)  -U\left(  r\right)  \right]  }},\label{Ueq}\\
\frac{d\phi}{d\rho}  &  =\frac{\ell/\rho^{2}}{\sqrt{2\left[  \mathcal{E}%
-\ell^{2}/\left(  2\rho^{2}\right)  -V\left(  \rho\right)  \right]  }}.
\label{Veq}%
\end{align}

\textit{Two three-dimensional central potentials, }$U\left(  r\right)
$\textit{ and }$V\left(  \rho\right)  $\textit{, are dual to each other, if }%
\begin{equation}
\frac{r^{2}}{l^{2}}\left[  E-U\left(  r\right)  \right]  =\frac{\rho^{2}}%
{\ell^{2}}\left[  \mathcal{E-}V\left(  \rho\right)  \right]  \label{rUrhoV}%
\end{equation}
\textit{with}%
\begin{equation}
r\leftrightarrow\rho^{\frac{l}{\ell}\sigma}. \label{rrho}%
\end{equation}
\textit{The orbit of the potential }$U\left(  r\right)  $\textit{ with the
energy }$E$\textit{ and the orbit of its dual potential }$V\left(
\rho\right)  $\textit{ with the energy }$\mathcal{E}$\textit{\ can be obtained
from each other by the replacement}%
\begin{equation}
\theta\leftrightarrow\frac{l}{\ell}\sigma\phi. \label{thetaphi}%
\end{equation}
\textit{Here }$\sigma$\textit{ is a constant chosen arbitrarily.}

\begin{proof}
By Eqs. (\ref{Ueq}) and (\ref{Veq}) we have%
\begin{align}
E-U\left(  r\right)   &  =\frac{1}{2}\frac{l^{2}}{r^{4}}\left(  \frac
{dr}{d\theta}\right)  ^{2}+\frac{l^{2}}{2r^{2}},\\
\mathcal{E-}V\left(  \rho\right)   &  =\frac{1}{2}\frac{\ell^{2}}{\rho^{4}%
}\left(  \frac{d\rho}{d\phi}\right)  ^{2}+\frac{\ell^{2}}{2\rho^{2}},
\end{align}
Substituting into Eq. (\ref{rUrhoV}) gives
\begin{equation}
\frac{r^{2}}{l^{2}}\left[  \frac{1}{2}\frac{l^{2}}{r^{4}}\left(  \frac
{dr}{d\theta}\right)  ^{2}+\frac{l^{2}}{2r^{2}}\right]  =\frac{\rho^{2}}%
{\ell^{2}}\left[  \frac{1}{2}\frac{\ell^{2}}{\rho^{4}}\left(  \frac{d\rho
}{d\phi}\right)  ^{2}+\frac{\ell^{2}}{2\rho^{2}}\right]  .
\end{equation}
This gives%
\begin{equation}
\left(  \frac{d\ln r}{d\theta}\right)  ^{2}=\left(  \frac{d\ln\rho}{d\phi
}\right)  ^{2},
\end{equation}
\qquad or,%
\begin{equation}
\frac{d\theta}{d\phi}=\frac{d\ln r}{d\ln\rho}.
\end{equation}
Because $\theta$ and $\phi$ are independent of $r$ and $\rho$, we have%
\begin{equation}
\frac{d\theta}{d\phi}=\frac{d\ln r}{d\ln\rho}=\sigma\frac{l}{\ell},
\label{Eqthetaphi}%
\end{equation}
where $\sigma$ is an arbitrary constant. Solving Eq. (\ref{Eqthetaphi}) gives
the duality transformations (\ref{rrho}) and (\ref{thetaphi}).
\end{proof}

\subsection{Power potential: Example}

The duality of a power potential, generally speaking, is no longer a power
potential. However, if we require that the duality is still a power potential,
we achieve the following result.

The dual potential of the power potential
\begin{equation}
U\left(  r\right)  =\xi r^{a},
\end{equation}
by the duality relations (\ref{rUrhoV}) and (\ref{rrho}), is%
\[
V\left(  \rho\right)  =\mathcal{-}\frac{\ell^{2}}{l^{2}}E\rho^{\frac{2l\sigma
}{\ell}-2}+\frac{\ell^{2}}{l^{2}}\xi\rho^{\frac{\left(  2+a\right)  l\sigma
}{\ell}-2}+\mathcal{E}%
\]%
\[
\sigma=\frac{2\ell}{\left(  2+a\right)  l}%
\]
If requiring that the dual potential $V\left(  \rho\right)  $ is still a power
potential, we can choose $\rho^{\frac{\left(  2+a\right)  l\sigma}{\ell}-2}%
=1$, which requires $\sigma=\frac{2\ell}{\left(  2+a\right)  l}$. Thus
$V\left(  \rho\right)  $ becomes%
\begin{equation}
V\left(  \rho\right)  =-\frac{\ell^{2}}{l^{2}}E\rho^{-\frac{2a}{a+2}}%
+\frac{\ell^{2}}{l^{2}}\xi+\mathcal{E}.
\end{equation}
Choosing $\mathcal{E}=-\frac{\ell^{2}}{l^{2}}\xi$ and $\eta=-\frac{\ell^{2}%
}{l^{2}}E$ gives a power potential%
\begin{equation}
V\left(  \rho\right)  =\eta r^{A}%
\end{equation}
with $\eta$ the coupling constant,
\begin{equation}
\frac{a+2}{2}=\frac{2}{A+2},
\end{equation}
and
\begin{align}
r  &  \leftrightarrow\rho^{\frac{2}{\left(  2+a\right)  }},\\
\theta &  \leftrightarrow\frac{2}{\left(  2+a\right)  }\phi.
\end{align}

\textit{Coulomb potential and harmonic-oscillator potential.} A special case
of the duality between the Coulomb potential and the harmonic-oscillator
potential is just the Newton-Hooke duality revealed by Newton. The Coulomb
potential $U\left(  r\right)  =\frac{\xi}{r}$, in fact, has an infinite number
of dual potentials corresponding to various choices of the parameter $\sigma$.
Choosing $\sigma=-2$, the dual relation (\ref{rUrhoV}) becomes $r^{2}\left(
\frac{\xi}{r}-E\right)  =\rho^{2}\left[  V\left(  \rho\right)  -\mathcal{E}%
\right]  $ with $r\leftrightarrow\rho^{2}$ and $\theta\leftrightarrow2\phi$.
The dual potential then reads $V\left(  \rho\right)  =-E\rho^{2}%
+\xi+\mathcal{E}$. Taking $\mathcal{E}=-\xi$, i.e., the energy of the system
becoming the coupling constant of its dual system, we arrive at a
harmonic-oscillator potential $V\left(  \rho\right)  =-E\rho^{2}=\eta\rho^{2}%
$. The energy of the Coulomb potential system becomes the coupling constant of
its dual potential. That is what Newton found.

\section{Duality in quantum mechanics \label{quantum}}

The equation of motion in quantum mechanics is the Schr\"{o}dinger equation.

\subsection{One-dimensional potential}

The one-dimensional stationary Schr\"{o}dinger equation $U\left(  x\right)  $
with the eigenvalue $E$ is%
\begin{equation}
\frac{d^{2}u\left(  x\right)  }{dx^{2}}+\left[  E-U\left(  x\right)  \right]
u\left(  x\right)  =0. \label{requ1d}%
\end{equation}

\textit{Two one-dimensional potentials, }$U\left(  x\right)  $\textit{ and
}$V\left(  \xi\right)  $\textit{, are dual to each other, if }%
\begin{equation}
\sigma\left\{  x^{2}\left[  U\left(  x\right)  -E\right]  +\frac{1}%
{4}\right\}  =\frac{1}{\sigma}\left\{  \xi^{2}\left[  V\left(  \xi\right)
-\mathcal{E}\right]  +\frac{1}{4}\right\}  . \label{rUrhoVq1d}%
\end{equation}
\textit{with}%
\begin{equation}
x\leftrightarrow\xi^{\sigma}. \label{rrhoq1d}%
\end{equation}
\textit{The eigenfunction of the potential }$U\left(  x\right)  $\textit{ and
the eigenfunction of its dual potential }$V\left(  \xi\right)  $\textit{ are
related by the replacement}%
\begin{equation}
u\left(  x\right)  \leftrightarrow\xi^{\left(  \sigma-1\right)  /2}v\left(
\xi\right)  . \label{uvq1d}%
\end{equation}
\textit{Here }$\sigma$\textit{ is a constant chosen arbitrarily.}

\begin{proof}
Performing the dual transforms (\ref{rrhoq1d}) and (\ref{uvq1d}) to the
one-dimensional stationary Schr\"{o}dinger equation Eq. (\ref{requ1d}) gives%
\begin{equation}
\frac{d^{2}v\left(  \xi\right)  }{d\xi^{2}}+\sigma^{2}\left\{  -\frac
{1-\sigma^{2}}{4\xi^{2}}-\xi^{2\left(  \sigma-1\right)  }\left[  U\left(
\xi^{\sigma}\right)  -E\right]  \right\}  v\left(  \xi\right)  =0.
\end{equation}
This is also a stationary Schr\"{o}dinger equation
\begin{equation}
\frac{d^{2}v\left(  \xi\right)  }{d\xi^{2}}+\left[  \mathcal{E}-V\left(
\xi\right)  \right]  v\left(  \xi\right)  =0
\end{equation}
with
\begin{equation}
V\left(  \xi\right)  =\sigma^{2}\left\{  \frac{1-\sigma^{2}}{4\xi^{2}}%
+\xi^{2\left(  \sigma-1\right)  }\left[  U\left(  \xi^{\sigma}\right)
-E\right]  \right\}  +\mathcal{E}\text{.}%
\end{equation}
This is just the dual relations (\ref{rUrhoVq1d}) and (\ref{rrhoq1d}).
\end{proof}

\subsection{Poschl-Teller potential: Example}

For the Poschl-Teller potential%
\begin{equation}
U\left(  x\right)  =\alpha\operatorname*{sech}\nolimits^{2}x,
\end{equation}
the stationary Schr\"{o}dinger equation has the following solution:%
\begin{equation}
u\left(  x\right)  =P_{\left(  \sqrt{1-4\alpha}-1\right)  /2}^{i\sqrt{E}%
}\left(  \tanh x\right)
\end{equation}
with $P_{l}^{m}\left(  z\right)  $ the associated Legendre polynomial.

The dual relations (\ref{rUrhoVq1d}) and (\ref{rrhoq1d}) give the dual
potential of the\textit{ }Poschl-Teller potential%
\begin{equation}
V\left(  \xi\right)  =\frac{1}{4}\left[  \sigma^{2}-1\right]  \frac{1}{\xi
^{2}}+\sigma^{2}\xi^{2\sigma-2}\left[  \alpha\operatorname*{sech}%
\nolimits^{2}\xi^{\sigma}-E\right]  +\mathcal{E} \label{PTV}%
\end{equation}
and its solution%
\begin{equation}
v\left(  \xi\right)  =\xi^{\frac{1-\sigma}{2}}P_{\left(  \sqrt{1-4\alpha
}-1\right)  /2}^{i\sqrt{E}}\left(  \tanh\xi^{\sigma}\right)  .
\end{equation}

Different choices of $\sigma$ give different dual potentials. The constant
$\mathcal{E}$ in the dual potential (\ref{PTV}) can also be chosen
arbitrarily, since it is a constant added in the potential.

Choosing $\sigma=1$ gives the Poschl-Teller potential itself. Concretely,
$\sigma=1$ gives $V\left(  \xi\right)  =\alpha\operatorname*{sech}%
\nolimits^{2}\xi-E+\mathcal{E}$. Taking $E=\mathcal{E}$ recovers the
Poschl-Teller potential.

Choosing $\sigma=\frac{1}{2}$ gives
\begin{equation}
V\left(  \xi\right)  =-\frac{3}{16\xi^{2}}+\frac{1}{4\xi}\left(
\alpha\operatorname*{sech}\nolimits^{2}\sqrt{\xi}+\alpha\xi-E\right)
-\frac{\alpha}{4}+\mathcal{E}.
\end{equation}
Taking $\mathcal{E}=\alpha/4$, i.e., letting the coupling constant in the
Poschl-Teller potential, $\alpha$, to be the energy eigenvalue of its dual
potential, gives the dual potential%
\begin{equation}
V\left(  \xi\right)  =-\frac{3}{16\xi^{2}}+\frac{1}{4\xi}\left(
\alpha\operatorname*{sech}\nolimits^{2}\sqrt{\xi}+\alpha\xi-E\right)  .
\end{equation}
The eigenfunction of the potential $V\left(  \xi\right)  $ with\ the
eigenvalue $\frac{\alpha}{4}$ then reads
\begin{equation}
v\left(  \xi\right)  =\xi^{1/4}P_{\left(  \sqrt{1-4\alpha}-1\right)
/2}^{i\sqrt{E}}\left(  \tanh\sqrt{\xi}\right)  .
\end{equation}

Choosing $\sigma=\frac{1}{3}$ gives%
\begin{equation}
V\left(  \xi\right)  =-\frac{2}{9\xi^{2}}+\frac{1}{9\xi^{4/3}}\left(
\alpha\operatorname*{sech}\nolimits^{2}\xi^{1/3}-\frac{2\alpha}{3}\xi
^{4/3}-E\right)  +\frac{2\alpha}{27}+\mathcal{E}.
\end{equation}
Taking $\mathcal{E}=-\frac{2\alpha}{27}$ gives the dual potential%
\begin{equation}
V\left(  \xi\right)  =-\frac{2}{9\xi^{2}}+\frac{1}{9\xi^{4/3}}\left(
\alpha\operatorname*{sech}\nolimits^{2}\xi^{1/3}-\frac{2\alpha}{3}\xi
^{4/3}-E\right)  .
\end{equation}
The eigenfunction of the potential $V\left(  \xi\right)  $ with\ the
eigenvalue $-\frac{2\alpha}{27}$ then reads%
\begin{equation}
v\left(  \xi\right)  =\xi^{1/3}P_{\left(  \sqrt{1-4\alpha}-1\right)
/2}^{i\sqrt{E}}\left(  \tanh\xi^{1/3}\right)  .
\end{equation}

\subsection{Three-dimensional central potential \label{3Dquantum}}

Two radial equations with potentials $U\left(  r\right)  $ and $V\left(
\rho\right)  $,%
\begin{align}
\frac{d^{2}u_{l}\left(  r\right)  }{dr^{2}}+\left[  E-\frac{l\left(
l+1\right)  }{r^{2}}-U\left(  r\right)  \right]  u_{l}\left(  r\right)   &
=0,\label{requ}\\
\frac{d^{2}v_{\ell}\left(  \rho\right)  }{d\rho^{2}}+\left[  \mathcal{E}%
-\frac{\ell\left(  \ell+1\right)  }{\rho^{2}}-V\left(  \rho\right)  \right]
v_{\ell}\left(  \rho\right)   &  =0.
\end{align}
which has the following duality relation.

\textit{Two central potentials, }$U\left(  r\right)  $\textit{ and }$V\left(
\rho\right)  $\textit{, are dual to each other, if}%
\begin{equation}
\frac{r^{2}}{\left(  l+\frac{1}{2}\right)  ^{2}}\left[  U\left(  r\right)
-E\right]  =\frac{\rho^{2}}{\left(  \ell+\frac{1}{2}\right)  ^{2}}\left[
V\left(  \rho\right)  -\mathcal{E}\right]  \label{rUrhoVq}%
\end{equation}
\textit{with}%
\begin{equation}
r\leftrightarrow\rho^{\sigma}. \label{rrhoq}%
\end{equation}
\textit{The radial eigenfunction of the potential }$U\left(  r\right)
$\textit{ and the radial eigenfunction of its dual potential }$V\left(
\rho\right)  $\textit{ are related by the replacement}%
\begin{equation}
u_{l}\left(  r\right)  \leftrightarrow\rho^{\left(  \sigma-1\right)
/2}v_{\ell}\left(  \rho\right)  . \label{uvq}%
\end{equation}
\textit{The relation between the angular momenta of the dual systems, then, is
}%
\begin{equation}
l+\frac{1}{2}\leftrightarrow\frac{1}{\sigma}\left(  \ell+\frac{1}{2}\right)  .
\label{ltrans}%
\end{equation}
\textit{Here }$\sigma$\textit{ is a constant chosen arbitrarily.}

\begin{proof}
The dual relations (\ref{rrhoq}) and (\ref{uvq}) transform Eq. (\ref{requ})
to
\begin{equation}
\frac{d^{2}v_{\ell}\left(  \rho\right)  }{d\rho^{2}}+\left\{  -\frac{\left[
\sigma\left(  l+\frac{1}{2}\right)  -\frac{1}{2}\right]  \left[  \sigma\left(
l+\frac{1}{2}\right)  +\frac{1}{2}\right]  }{\rho^{2}}-\sigma^{2}%
\rho^{-2\left(  1-\sigma\right)  }\left[  U\left(  \rho^{\sigma}\right)
-E\right]  \right\}  v_{\ell}\left(  \rho\right)  =0. \label{requq}%
\end{equation}
Eq. (\ref{requq}) is the eigenequation of the potential%
\begin{equation}
V\left(  \rho\right)  =\sigma^{2}\rho^{-2\left(  1-\sigma\right)  }\left[
U\left(  \rho^{\sigma}\right)  -E\right]  +\mathcal{E} \label{VrhoUr}%
\end{equation}
with the energy $\mathcal{E}$ and the angular momentum%
\begin{equation}
\ell\left(  \ell+1\right)  =\left[  \sigma\left(  l+\frac{1}{2}\right)
-\frac{1}{2}\right]  \left[  \sigma\left(  l+\frac{1}{2}\right)  +\frac{1}%
{2}\right]  , \label{angularm}%
\end{equation}
i.e.,%
\begin{equation}
\frac{d^{2}v_{\ell}\left(  \rho\right)  }{d\rho^{2}}+\left[  \mathcal{E}%
-\frac{\ell\left(  \ell+1\right)  }{\rho^{2}}-V\left(  \rho\right)  \right]
v_{\ell}\left(  \rho\right)  =0.
\end{equation}
Eq. (\ref{VrhoUr}) gives the duality relation (\ref{rUrhoVq}) and Eq.
(\ref{angularm}) gives the duality relations (\ref{ltrans}) and (\ref{rrhoq}).
\end{proof}

It should be emphasized that the dual relation of the angular momentum, Eq.
(\ref{ltrans}), is a result of the dual transforms (\ref{rrhoq}) and
(\ref{uvq}).

\subsection{Power potential: Example}

Like that in classical mechanics, in quantum mechanics a potential can also
have an arbitrary number of dual potentials. The dual potential of a power
potential, generally speaking, is not a power potential. Nevertheless, if we
still require that the dual potential of a power potential%
\begin{equation}
U\left(  r\right)  =\xi r^{a}%
\end{equation}
is also a power potential, we can choose $\sigma=\frac{2}{2+a}$\ in Eq.
(\ref{rrhoq}). Then the dual potential reads%
\begin{equation}
V\left(  \rho\right)  =-\sigma^{2}E\rho^{-\frac{2a}{a+2}}+\sigma^{2}%
\xi+\mathcal{E}.
\end{equation}
Choosing%
\begin{align}
\mathcal{E}  &  =-\sigma^{2}\xi,\\
\eta &  =-\sigma^{2}E,
\end{align}
we arrive at a power potential%
\begin{equation}
V\left(  \rho\right)  =\eta r^{A}.
\end{equation}
The corresponding dual relations are%
\begin{equation}
\frac{a+2}{2}=\frac{2}{A+2} \label{aA}%
\end{equation}
and
\begin{align}
r  &  \leftrightarrow\rho^{2/\left(  a+2\right)  },\label{rrhopower}\\
u_{l}\left(  r\right)   &  \leftrightarrow\rho^{-\frac{a}{2\left(  a+2\right)
}}v_{\ell}\left(  \rho\right)  . \label{radialwavepower}%
\end{align}

\textit{Coulomb potential and harmonic-oscillator potential. }Like that in the
classical mechanics, the dual potential of the Coulomb potential $U\left(
r\right)  =\frac{\xi}{r}$ with $\sigma=-2$ is the harmonic-oscillator
potential. The dual transforms (\ref{aA}), (\ref{rrhopower}), and
(\ref{radialwavepower}) become%
\begin{align}
r  &  \leftrightarrow\rho^{2},\label{rcoul}\\
u_{l}\left(  r\right)   &  \leftrightarrow\rho^{1/2}v_{\ell}\left(
\rho\right)  , \label{ucoul}%
\end{align}
and%
\begin{equation}
l+\frac{1}{2}\leftrightarrow\frac{1}{2}\left(  \ell+\frac{1}{2}\right)  .
\label{lcoul}%
\end{equation}
The dual relation (\ref{rUrhoVq}) now becomes $2r^{2}\left[  U\left(
r\right)  -E\right]  =\frac{1}{2}\rho^{2}\left[  V\left(  \rho\right)
-\mathcal{E}\right]  $. We have%
\begin{equation}
V\left(  \rho\right)  =-4E\rho^{2}+4\xi+\mathcal{E}.
\end{equation}
Taking $\mathcal{E}=-4\xi$ gives%
\begin{equation}
V\left(  \rho\right)  =-4E\rho^{2}.
\end{equation}

The radial eigenfunction of the Coulomb potential is%
\begin{equation}
u_{l}\left(  r\right)  =A_{l}e^{-\sqrt{-E}r}\left(  2\sqrt{-E}\right)
^{l+1}r^{l+1}\text{ }_{1}F_{1}\left(  l+1+\frac{\xi}{2\sqrt{-E}},2\left(
l+1\right)  ,2\sqrt{-E}r\right)  .
\end{equation}
The dual transforms (\ref{rcoul}), (\ref{ucoul}), and (\ref{lcoul}) give the
radial eigenfunction of the Coulomb potential:%
\begin{equation}
v_{\ell}\left(  \rho\right)  =A_{l}e^{-\sqrt{-E}\rho^{2}}\left(  2\sqrt
{-E}\right)  ^{\frac{\ell}{2}+\frac{3}{4}}\rho^{\ell+1}\text{ }_{1}%
F_{1}\left(  \frac{\ell}{2}+\frac{3}{4}+\frac{\xi}{2\sqrt{-E}},\ell+\frac
{3}{2},2\sqrt{-E}\rho^{2}\right)  .
\end{equation}
This is the solution of the radial equation of the central\textit{
}harmonic-oscillator potential with the eigenvalue $\mathcal{E}=-4\xi$ and the
coupling constant $\eta=-4E$:%
\begin{equation}
\frac{d^{2}v_{\ell}\left(  \rho\right)  }{d\rho^{2}}+\left[  -4\xi-\frac
{\ell\left(  \ell+1\right)  }{\rho^{2}}+4E\rho^{2}\right]  v_{\ell}\left(
\rho\right)  =0.
\end{equation}

\section{Duality family}

A mechanical system can have an infinite number of dual systems, since there
exist infinite choices of the parameter $\sigma$ in, e.g., the dual relation
(\ref{rrho1d}) and (\ref{thetaphi1d}). All dual mechanical systems form a
duality family consisting of infinite members. All members in a duality family
are dual to each other. This implies that once the solution of a mechanical
system in the duality family is obtained, all other mechanical systems in the
duality family are obtained immediately through the duality transform.

Inspection of the duality relation shows that there exist algebraic structures
in the duality family.

\section{Conclusion \label{Conclusion}}

A duality in classical and quantum mechanics is revealed. In classical
mechanics the dynamical equation is the Newton equation; in quantum mechanics
the dynamical equation is the Schr\"{o}dinger equation. In future works, we
will discuss the duality of the potential in relativistic quantum mechanics:
the duality of the Klein-Gordon equation and the duality of the Dirac
equation. Moreover, in the present paper we only consider the duality of
conservative forces which can be described by a potential. We will take the
nonconservative force, such as the magnetic force, into account. In previous
works, we provide some examples of such a duality \cite{li2017quantum}.
Moreover, it is also shown that there exists similar dualities in
self-interacting scalar fields \cite{li2019duality1,li2019duality2} and in the
Gross-Pitaevskii equation \cite{liu2021exactly}.

In the subsequent work, we will concentrate on the duality family. In such a
duality family, all potentials are dual to each other and the solutions of the
potentials in the family can be transformed to each other by the dual
relation. It is worthy to discuss the general properties of the duality family.

\acknowledgments

We are very indebted to Dr G. Zeitrauman for his encouragement. This work is supported in part by Special Funds for theoretical physics Research Program of the NSFC under Grant No. 11947124, and NSFC under Grant Nos. 11575125 and 11675119.






\end{CJK*}
\end{document}